\documentclass[12pt]{article}
\usepackage{graphicx}
\usepackage{endfloat}

\begin{document}
\title{ Synthesizing NMR analogues of Einstein-Podolsky-Rosen states using
 generalized Grover's algorithm}
\author{\small{} Jingfu Zhang$^{1}$ , Zhiheng Lu$^{1}$,Lu Shan$^{2}$,and Zhiwei
Deng$^{2}$  \\
\small{} $^{1}$Department of Physics,\\
\small{}Beijing Normal University, Beijing,
100875, Peoples' Republic of China\\
\small{} $^{2}$Testing and Analytical Center,\\
\small{}  Beijing Normal University,
 Beijing,100875, Peoples' Republic of China}
\date{}
\maketitle
\begin{minipage}{120mm}
\hspace{0.5cm} {\small
  By designing a proper unitary
operator U, we synthesize NMR analogues of Einstein-Podolsky-Rosen
states (pseudo-EPR states) using generalized Grover's algorithm on
a nuclear magnetic resonance (NMR) quantum computer. Experiments
also demonstrate generalized Grover's algorithm for the case in
which there are multiple marked states.}

 PACS number(s):03.67
\end{minipage}
\vspace{0.3cm}

  Since the quantum searching algorithm was first
proposed by Grover [1], several generalizations of the original
algorithm have been developed [2]-[4]. One of the generalized
algorithms can be posed as follows. For a system with a total of
$N$ basis states, a composite operator $Q$ is defined as
$Q\equiv-I_{s}U^{-1}I_{t}U$. $U$ can be almost any valid quantum
mechanical unitary operator. $I_{s}$ is defined as $I_{s}\equiv
I-2|s><s|$, where $I$ denotes unit matrix. $|s>$ denotes a
predefined basis state that is used as an initial state in our
experiments. $I_{s}$ is a diagonal matrix with all diagonal
elements equal to 1 except the $ss$th elements which are -1.
Similarly, $I_{t}$ can be written as $I_{t}\equiv
I-\sum_{t}2|t><t|$, where $|t>$ denote the marked states, and
there are $r$ marked states. For any $|t>$, $I_{t}|t>=-|t>$.  $u$
is defined as $u=\sqrt{\sum_{t}|U_{ts}|^{2}}$, where
$U_{ts}=<t|U|s>$. It has been proved that $\pi/4u$ applications of
$Q$ transform the system from $|s>$ into a superposition denoted
as $|\psi>=\frac{1}{u}\sum_{t}U_{ts}U^{-1}|t>$ if $u\ll 1$.
Through introducing an ancilla qubit and choosing a proper $U$,
Grover proposed a theoretical scheme to synthesize a specified
quantum superposition on $N$ states in $O(\sqrt{N})$ steps using
the generalized algorithm [5]. Nevertheless, we find that some
useful superpositions, such as NMR analogues of
Einstein-Podolsky-Rosen states (pseudo-EPR states) [6]-[8], can be
synthesized using the algorithm without an ancilla qubit. Such
superpositions can be represented as
$|\psi_{su}>=\frac{1}{\sqrt{r}}\sum_{t}e^{i\phi_{t}}|t>$ (su for
superposition), where $\phi_{t}$ denote the phases of $|t>$. By
designing a proper $U$, we make $|U_{ts}|$ identical, and let
$\frac{U_{ts}}{|U_{ts}|}=e^{i\phi_{t}}$, so that $|\psi>$ can be
represented as
$|\psi>=\frac{1}{\sqrt{r}}\sum_{t}e^{i\phi_{t}}U^{-1}|t>$. After
the application of $U$, the system lies in $|\psi_{su}>$, where an
irrelevant overall phase factor can be ignored.

   In our previous work, we have realized generalized Grover's
searching algorithm for the case in which there is one marked
state on a two-qubit NMR quantum computer [9]. In this paper, we
will synthesize pseudo-EPR states using the algorithm.

   Our experiments use a sample of carbon-13 labelled chloroform
dissolved in d6-acetone. Data are taken at room temperature with a
Bruker DRX 500 MHz spectrometer. The resonance frequencies
$\nu_{1}=125.76$ MHz for $^{13}C$, and $\nu_{2}=500.13$ MHz for
$^{1}H$. The coupling constant $J$ is measured to be 215 Hz. If
the magnetic field is along $\hat{z}$-axis, by setting $\hbar=1$,
the Hamitonian of this system is represented as
\begin{equation}\label{1}
  H=-2\pi\nu_{1}I_{z}^{1}-2\pi\nu_{2}I_{z}^{2}+2\pi J I_{z}^{1}
  I_{z}^{2},
\end{equation}
where $I_{z}^{k}(k=1,2)$ are the matrices for $\hat{z}$-component
of the angular momentum of the spins [10]. In the rotating frame
of spin $k$, the evolution caused by a radio-frequency(rf) pulse
on resonance along $\hat{x}$ or $-\hat{y}$-axis is denoted as $
X_{k}(\varphi_{k})= e^{i\varphi_{k}I_{x}^{k}}$ or $
Y_{k}(-\varphi_{k})= e^{-i\varphi_{k}I_{y}^{k}}$, where
$\varphi_{k}=B_{1}\gamma_{k}t_{p}$ with $k$ specifying the
affected spin. $B_{1} $, $\gamma_{k}$ and $t_{p}$ represent the
strength of rf pulse, gyromagnetic ratio and the width of rf
pulse, respectively. The pulse used above is denoted as
$[\varphi]_{x}^{k}$ or $[-\varphi]_{y}^{k}$.
 The coupled-spin evolution is denoted as
\begin{equation}\label{2}
  [\tau]=e^{-i2\pi J I_{z}^{1} I_{z}^{2}\tau},
\end{equation}
where $\tau$ is evolution time. The initial pseudo-pure state
\begin{equation}\label{3}
  |s>=|\uparrow>_{1}|\uparrow>_{2}=\left(\begin{array}{c}
    1 \\
    0 \\
    0 \\
    0 \
  \end{array}\right)
\end{equation}
is prepared by using spatial averaging [11], where
$|\uparrow>_{k}$ denotes the state of spin $k$. For convenience,
the notation $|\uparrow>_{1}|\uparrow>_{2}$ is simplified as
$|\uparrow\uparrow>$. The basis states are arrayed as
$|\uparrow\uparrow>,|\uparrow\downarrow>,|\downarrow\uparrow>$, $
|\downarrow\downarrow>$. Pseudo-EPR states are denoted as
$|\psi_{1}>=(|\uparrow\uparrow>+|\downarrow\downarrow>)/\sqrt{2}$,
$|\psi_{2}>=(|\uparrow\uparrow>-|\downarrow\downarrow>)/\sqrt{2}$,
$|\psi_{3}>=(|\uparrow\downarrow>+|\downarrow\uparrow>)/\sqrt{2}$,
and
$|\psi_{4}>=(|\uparrow\downarrow>-|\downarrow\uparrow>)/\sqrt{2}$.
EPR (or pseudo-EPR) states are very useful in quantum information
and have be implemented in experiments [12][13]. We will
synthesize pseudo-EPR states using generalized Grover's algorithm.

   $U$ is chosen as $U=Y_{1}(\varphi_{1})Y_{2}(\varphi_{2})$
represented as
\begin{equation}\label{4}
  U=\left(\begin{array}{cccc}
    c_{1}c_{2} & c_{1}s_{2} & s_{1}c_{2} &s_{1}s_{2} \\
    -c_{1}s_{2} &  c_{1}c_{2} & -s_{1}s_{2} & s_{1}c_{2} \\
    -s_{1}c_{2} & -s_{1}s_{2} & c_{1}c_{2} & c_{1}s_{2} \\
    s_{1}s_{2} & -s_{1}c_{2} & -c_{1}s_{2} & c_{1}c_{2} \
  \end{array}\right),
 \end{equation}
where $c_{k}\equiv cos(\varphi_{k}/2)$, $s_{k}\equiv
sin(\varphi_{k}/2)$. According to the first column of $U$, we
design $\varphi_{1}$ and $\varphi_{2}$ for synthesizing pseudo-EPR
states. $U$ is chosen as
$U_{1}=Y_{1}(\frac{\pi}{4})Y_{2}(\frac{3\pi}{4})$,
$U_{2}=Y_{1}(\frac{\pi}{4})Y_{2}(-\frac{3\pi}{4})$,
$U_{3}=Y_{1}(\frac{\pi}{4})Y_{2}(\frac{\pi}{4})$, and
$U_{4}=Y_{1}(-\frac{\pi}{4})Y_{2}(\frac{\pi}{4})$ for
$|\psi_{1}>$, $|\psi_{2}>$, $|\psi_{3}>$, and $|\psi_{4}>$,
respectively. One can prove that
$u_{j}=\sqrt{\sum_{t}|U_{jts}|^{2}}=1/2$, where j=1,2,3 or 4. The
following discussion will show the condition $u\ll 1$ is not
necessary. If the system starts with other basis states, $U$ is
chosen as other forms. For example, if
$I_{s}=|\uparrow\downarrow>$, according to the second column of
$U$, $U$ is chosen as
$U_{1}=Y_{1}(-\frac{\pi}{4})Y_{2}(\frac{\pi}{4})$,
$U_{2}=Y_{1}(\frac{\pi}{4})Y_{2}(\frac{\pi}{4})$,
$U_{3}=Y_{1}(\frac{\pi}{4})Y_{2}(-\frac{3\pi}{4})$, and
$U_{4}=Y_{1}(\frac{\pi}{4})Y_{2}(\frac{3\pi}{4})$. If $U$ is
chosen as $U_{1}=X_{1}(\frac{\pi}{4})Y_{2}(\frac{3\pi}{4})$,
$U_{2}=X_{1}(\frac{\pi}{4})Y_{2}(-\frac{3\pi}{4})$,
$U_{3}=X_{1}(\frac{\pi}{4})Y_{2}(\frac{\pi}{4})$, and
$U_{4}=X_{1}(\frac{\pi}{4})Y_{2}(-\frac{\pi}{4})$,
pseudo-entangled states
$(|\uparrow\uparrow>-i|\downarrow\downarrow>)/\sqrt{2}$,
$(|\uparrow\uparrow>+i|\downarrow\downarrow>)/\sqrt{2}$,
$(|\uparrow\downarrow>-i|\downarrow\uparrow>)/\sqrt{2}$, and
$(|\uparrow\downarrow>+i|\downarrow\uparrow>)/\sqrt{2}$ can be
obtained, respectively.

 If $|s>=|\uparrow\uparrow>$,
$I_{s}$ is chosen as $I_{0}$ represented as
\begin{equation}\label{5}
  I_{0}=\left(\begin{array}{cccc}
    -1 & 0 & 0 & 0 \\
    0 & 1 & 0 & 0 \\
    0 & 0 & 1 & 0 \\
    0 & 0 & 0 & 1 \
  \end{array}\right).
 \end{equation}
Because $r=N/2$, the conditional sign flip operators for
$|\uparrow\uparrow>$ and $|\downarrow\downarrow>$, and for
$|\uparrow\downarrow>$ and $|\downarrow\uparrow>$ can be chosen
the same form represented as
\begin{equation}\label{6}
  I_{t}=\left(\begin{array}{cccc}
    1 & 0 & 0 & 0 \\
    0 & -1 & 0 & 0 \\
    0 & 0 & -1 & 0 \\
    0 & 0 & 0 & 1 \
  \end{array}\right).
 \end{equation}
$Q$ is represented as $Q_{j}\equiv-I_{s}U_{j}^{-1}I_{t}U_{j}$ for
$|\psi_{j}>$. The operator $G_{j}^{(n)}$ is defined as
$G_{j}^{(n)}\equiv U_{j}Q_{j}^{n}$, which means that operation
$Q_{j}$ is repeated $n$ times, and then $U_{j}$ is applied. It is
easy to prove that $G_{1}^{(1)}|\uparrow\uparrow>=-|\psi_{1}>$,
$G_{2}^{(1)}|\uparrow\uparrow>=-|\psi_{2}>$,
$G_{3}^{(1)}|\uparrow\uparrow>=-|\psi_{3}>$, and
$G_{4}^{(1)}|\uparrow\uparrow>=-|\psi_{4}>$. The required $n$ to
synthesize a target state displays a period of 3. For example,
$G_{3}^{(4)}|\uparrow\uparrow>=|\psi_{3}>$,
$G_{3}^{(7)}|\uparrow\uparrow>=-|\psi_{3}>$.

   The following rf and gradient pulse sequence
$[\alpha]_{x}^{2}-[grad]_{z}-[\pi/4]_{x}^{1}-1/4J-[\pi]_{x}^{1,2}-1/4J-[-\pi]_{x}^{1,2}
-[-\pi/4]_{y}^{1}-[grad]_{z}$ transforms the system from the
equilibrium state represented as
\begin{equation}\label{7}
  \rho_{eq}=\gamma_{1} I_{z}^{1}+\gamma_{2} I_{z}^{2}
\end{equation}
to the initial state represented as
\begin{equation}\label{8}
  \rho_{0}=I_{z}^{1}/2+I_{z}^{2}/2+I_{z}^{1}I_{z}^{2},
\end{equation}
which can be used as the pseudo-pure state $|\uparrow\uparrow>$
[14]. $\alpha=\arccos(\gamma_{1}/2\gamma_{2})$,
 $[grad]_{z}$ denotes gradient pulse along $\hat{z}$-axis, and the
symbol 1/4J means that the system evolutes under the Hamitonian
$H$ for 1/4J time when pulses are switched off. The pulses are
applied from left to right. $[\pi]_{x}^{1,2}$ denotes a
nonselective pulse (hard pulse). The evolution caused by the pulse
sequence $1/4J-[\pi]_{x}^{1,2}-1/4J-[-\pi]_{x}^{1,2}$ is
equivalent to the coupled-spin evolution $[1/2J]$ described in
Eq.(2) [15]. $[\pi]_{x}^{1,2}$ pulses are applied in pairs each of
which take opposite phases in order to reduce the error
accumulation causes by imperfect calibration of $[\pi]$ pulses
[16].

$U_{1}$, $U_{2}$, $U_{3}$ and $U_{4}$ are realized by
$[\pi/4]_{y}^{1}-[3\pi/4]_{y}^{2}$,
$[\pi/4]_{y}^{1}-[-3\pi/4]_{y}^{2}$,
$[\pi/4]_{y}^{1}-[\pi/4]_{y}^{2}$, and
$[-\pi/4]_{y}^{1}-[\pi/4]_{y}^{2}$, respectively. $I_{t}=[1/J]$,
 realized by $1/2J-[\pi]_{x}^{1,2}-1/2J-[-\pi]_{x}^{1,2}$.
According to Ref.[17], $I_{0}$ is realized by
$1/4J-[\pi]_{x}^{1,2}-1/4J-[-\pi]_{x}^{1,2}-[-\pi/2]_{y}^{1,2}-[-\pi/2]_{x}^{1,2}-
[\pi/2]_{y}^{1,2}$. $G_{j}^{(1)}$ transforms the system from the
initial state into the corresponding target state. For example,
$G_{3}^{(1)}$ transforms the system from $\rho_{0}$ into
$\rho_{3}$ represented as
\begin{equation}\label{9}
  \rho_{3}=I_{x}^{1}I_{x}^{2}+I_{y}^{1}I_{y}^{2}-I_{z}^{1}I_{z}^{2},
\end{equation}
which is equivalent to $|\psi_{3}><\psi_{3}|$. A readout pulse
$[\pi/2]_{y}^{1}$ transforms $\rho_{3}$ to
$I_{z}^{1}I_{x}^{2}+I_{y}^{1}I_{y}^{2}+I_{x}^{1}I_{z}^{2}$, which
is equivalent to
\begin{equation}\label{10}
  \rho_{3r}=\frac{1}{4}\left(\begin{array}{cccc}
    1 & 1 & 1 & -1 \\
    1 & 1 & 1 & -1 \\
    1 & 1 & 1 & -1 \\
    -1 & -1 & -1 & 1 \
  \end{array}\right).
\end{equation}
The information on matrix elements (1,3) and (2,4) in Eq.(10) can
be directly obtained in the carbon spectrum, and the information
on elements (1,2) and (3,4) can be directly obtained in the proton
spectrum. Similarly, when the system lies in $|\psi_{1}>$,
$|\psi_{2}>$, or $|\psi_{4}>$, through the readout pulse
$[\pi/2]_{y}^{1}$, the system lies in the state described as
\begin{equation}\label{11}
  \rho_{1r}=\frac{1}{4}\left(\begin{array}{cccc}
    1 & 1 & -1 & 1 \\
    1 & 1 & -1 & 1 \\
    -1 & -1 & 1 & -1 \\
    1 & 1 & -1 & 1 \
  \end{array}\right),
\end{equation}

\begin{equation}\label{12}
  \rho_{2r}=\frac{1}{4}\left(\begin{array}{cccc}
    1 & -1 & -1 & -1 \\
    -1 & 1 & 1 & 1 \\
    -1 & 1 & 1 & 1 \\
    -1 & 1 & 1 & 1 \
  \end{array}\right),
\end{equation}
or
\begin{equation}\label{13}
  \rho_{4r}=\frac{1}{4}\left(\begin{array}{cccc}
    1 & -1 & 1 & 1\\
    -1 & 1 & -1 & -1 \\
    1 & -1 & 1 & 1 \\
    1 & -1 & 1 & 1 \
  \end{array}\right).
\end{equation}
Through observing the matrix elements (1,3), (2,4), (1,2) and
(3,4) in Eqs.(10)-(13), one can distinguish pseudo-EPR states.

  In experiments, for each target state, the carbon spectrum and
proton spectrum are recorded in two experiments. For different
target states, carbon spectra or proton spectra are recorded in an
identical fashion. Because the absolute phase of an NMR signal is
not meaningful, we must use reference signals to adjust carbon
spectra and proton spectra so that the phases of the signals are
meaningful [18]. When the system lies in the initial pseudo-pure
state described as Eq.(8), the readout pulses $[\pi/2]_{y}^{1}$
and $[\pi/2]_{y}^{2}$ transform it into states represented as
\begin{equation}\label{14}
  \rho_{sr1}=\frac{1}{4}\left(\begin{array}{cccc}
    1 & 0 & -2 & 0\\
    0 & -1 & 0 & 0 \\
    -2 & 0 & 1 & 0 \\
    0 & 0 & 0 & -1 \
  \end{array}\right),
\end{equation}
 and
\begin{equation}\label{15}
  \rho_{sr2}=\frac{1}{4}\left(\begin{array}{cccc}
    1 & -2 & 0 & 0\\
    -2 & 1 & 0 & 0 \\
    0 & 0 & -1 & 0 \\
    0 & 0 & 0 & -1 \
  \end{array}\right),
\end{equation}
respectively. In the carbon spectrum or proton spectrum, there is
only one NMR peak corresponding to element (1,3) in $\rho_{sr1}$
or to element (1,2) in $\rho_{sr2}$. Through calibrating the
phases of the two signals, the two peaks are adjusted into
absorption shapes which are shown as Fig.1(a) for carbon spectrum
and Fig.1(b) for proton spectrum. The two signals are used as
reference signals of which phases are recorded to calibrate the
phases of signals in other carbon spectra and proton spectra,
respectively. One should note that the minus elements in Eq.(14)
and Eq.(15) correspond to the positive peaks in Fig.1(a) and
Fig.1(b).

  Experiments start with pseudo-pure state $|\uparrow\uparrow>$. $G_{j}^{(1)}$
transforms $|\uparrow\uparrow>$ into $|\psi_{j}>$. If no readout
pulse is applied, the amplitudes of peaks is so small that they
can be ignored. By applying the spin-selective readout pulse
$[\pi/2]_{y}^{1}$, we obtain carbon spectra shown in Figs.2(a),
(b), (c), and (d), and proton spectra shown in Figs.2(e), (f),
(g), and (h), corresponding to $|\psi_{1}>$, $|\psi_{2}>$,
$|\psi_{3}>$, and $|\psi_{4}>$, respectively. In Fig.2(a), for
example, the right and left peaks correspond to the matrix
elements (1,3) and (2,4) in Eq.(11), respectively. Similarly, in
Fig.2(e), the two peaks correspond to the matrix elements (1,2)
and (3,4) in Eq.(11). The phases of the signals corroborate the
synthesis of pseudo-EPR states.

   In conclusion, we synthesize pseudo-EPR states
using the generalized Grover's algorithm by choosing a proper $U$.
Although the ancilla qubit is not used, our experimental scheme
shows the essential meaning of Grover's original idea. The
experiments also demonstrate generalized Grover's algorithm for
the case in which there are $N/2$ marked states.

  This work was partly supported by the National Nature Science
Foundation of China. We are also grateful to Professor Shouyong
Pei of Beijing Normal University for his helpful discussions on
the principle of quantum algorithm.
\newpage
\bibliographystyle{article}

\newpage
{\begin{center}\large{Figure Captions}\end{center}
\begin{enumerate}
\item The carbon spectrum (Fig.1(a)) obtained through selective readout
pulse for $^{13}C$ $[\pi/2]_{y}^{1}$  and the proton spectrum
(Fig.2(b)) obtained through selective readout pulse for $^{1}H$
$[\pi/2]_{y}^{2}$ when the two-spin system lies in pseudo-pure
state $|\uparrow\uparrow>$. The amplitude has arbitrary units. The
two peaks are adjusted into absorption shapes. The two signals are
used as reference signals to adjust other spectra.
\item Carbon spectra (shown by the left column) and
proton spectra (shown by the right column)
 obtained through $[\pi/2]_{y}^{1}$ after pseudo-EPR
 states are synthesized. Figs.2(a)- (d) and Figs.2(e)- (h) correspond to states
$(|\uparrow\uparrow>+|\downarrow\downarrow>)/\sqrt{2}$,
 $(|\uparrow\uparrow>-|\downarrow\downarrow>)/\sqrt{2}$,
$(|\uparrow\downarrow>+|\downarrow\uparrow>)/\sqrt{2}$, and
$(|\uparrow\downarrow>-|\downarrow\uparrow>)/\sqrt{2}$,
respectively.
\end{enumerate}
\begin{figure}{1}
\includegraphics[width=0.75in]{Fig.1.eps}
\caption{}
\end{figure}
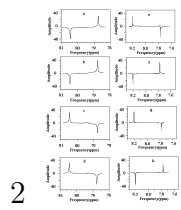
\begin{figure}{2}
\includegraphics[width=0.75in]{Fig.2.eps}
\caption{}
\end{figure}

\end{document}